\documentstyle[seceq,epsf]{ptptex}
\newcommand{\gsim}{\raisebox{-0.55ex}{$\stackrel{\displaystyle >}{\sim}$}}
\newcommand{\diff}[2]{\frac{\partial #1}{\partial #2}}

\notypesetlogo  %comment in if to eliminate PTPTeX logo
\title{%        %You can use \\ for explicit line-break
Scaling Behavior of $(N_{\rm ch})^{-1}dN_{\rm ch}/d\eta$ at $\sqrt{s_{NN}} = 130\ {\rm GeV}$ by the PHOBOS Collaboration and Its Implication
}
\subtitle{A Possible Explanation Employing the Ornstein-Uhlenbeck Process}    %use this when you want a subtitle

\author{%       %Use \sc for the family name
Minoru Biyajima$^1$, Masaru Ide$^1$, Takuya Mizoguchi$^2$ and Naomichi Suzuki$^3$
}

\inst{%         %Affiliation, neglected when [addenda] or [errata]
$^1$Department of Physics, Faculty of Science, Shinshu University, Matsumoto 390-8621, Japan\\
$^2$Toba National College of Maritime Technology, Toba 517-8501, Japan\\
$^3$Matsusho Gakuen Junior College, Matsumoto 390-1295, Japan}

\recdate{%      %Editorial Office will fill in this.
March 22, 2002
}

\abst{%       %this abstract is neglected when [addenda] or [errata]
Recently, interesting data concerning $dN_{\rm ch}/d\eta$ in Au-Au collisions [$\eta=-\ln \tan (\theta/2)$] with centrality cuts have been reported from the PHOBOS Collaboration. In most treatment these data are divided by the number of participants (nucleons) in collisions. Instead of this method, we use the total multiplicity $N_{\rm ch} = \int (dN_{\rm ch}/d\eta)d\eta$ and find that there is scaling phenomenon among $(N_{\rm ch})^{-1}dN_{\rm ch}/d\eta = dn/d\eta$ with different centrality cuts at $\sqrt{s_{NN}} = 130$ GeV. To explain this scaling behavior of $dn/d\eta$, we employ a stochastic approach using the Ornstein-Uhlenbeck process with two sources. A Langevin equation is adopted for this explanation. Moreover, comparisons of $dn/d\eta$ at $\sqrt{s_{NN}} = 130$ GeV with that at $\sqrt{s_{NN}} = 200$ GeV are made, and no significant difference is found. A possible method fot the detection of the quark-gluon plasma (QGP) through $dN_{\rm ch}/d\eta$ is presented.
}

\begin{document}

\maketitle

%   SECTION 1
%
\section{Introduction}
Recently, interesting data from the PHOBOS Collaboration on $dN_{\rm ch}/d\eta$ [$\eta=-\ln \tan (\theta/2)$]\footnote{
Here, 
\begin{eqnarray*}
  y &=& \frac 12 \ln \frac{E+p_z}{E-p_z} = \frac 12 \ln\left[\frac{\sqrt{1+  m^2/p_t^2+\sinh^2 \eta} + \sinh \eta}{\sqrt{1+ m^2/p_t^2+\sinh^2 \eta} - \sinh \eta}\right] = \tanh^{-1} \left(\frac{p_z}E\right) \approx -\ln\tan(\theta/2) \equiv \eta\, .\\
  \eta &=& \frac 12 \ln \frac{p+p_z}{p-p_z}\ {\rm and}\quad \frac{dn}{d\eta} = \frac pE \frac{dn}{dy}\:,\ {\rm where }\quad \frac pE= \frac{\cosh \eta}{\sqrt{1+ m^2/p_t^2+\sinh^2 \eta}}\:.
\end{eqnarray*}
\label{foot1}
}in Au-Au collisions at $\sqrt{s_{NN}} = 130$ GeV have been published.\cite{Back:2001bq} The authors of Ref.~\citen{Back:2001bq} calculated the quantity
\begin{eqnarray}
\frac 1{\langle N_{\rm part}\rangle/2}\frac{dN_{\rm ch}}{d\eta} = f(\langle N_{\rm part}\rangle,\,N_{\rm coll},\,\eta)\:,
\label{eq1.1}
\end{eqnarray}
where $N_{\rm part}$ and $N_{\rm coll}$ represent the number of participants (nucleons) and the number of collision particles in Au-Au collisions. The quantity in Eq.~(\ref{eq1.1}) depends on the centrality cuts. The intercept $f(N_{\rm part},\,N_{\rm coll},\,\eta=0)$ is an increasing function of $\langle N_{\rm part}\rangle$.

\begin{figure}[htb]
  \epsfxsize= 10 cm   %or \epsfysize= HEIGHT cm
  \centerline{\epsfbox{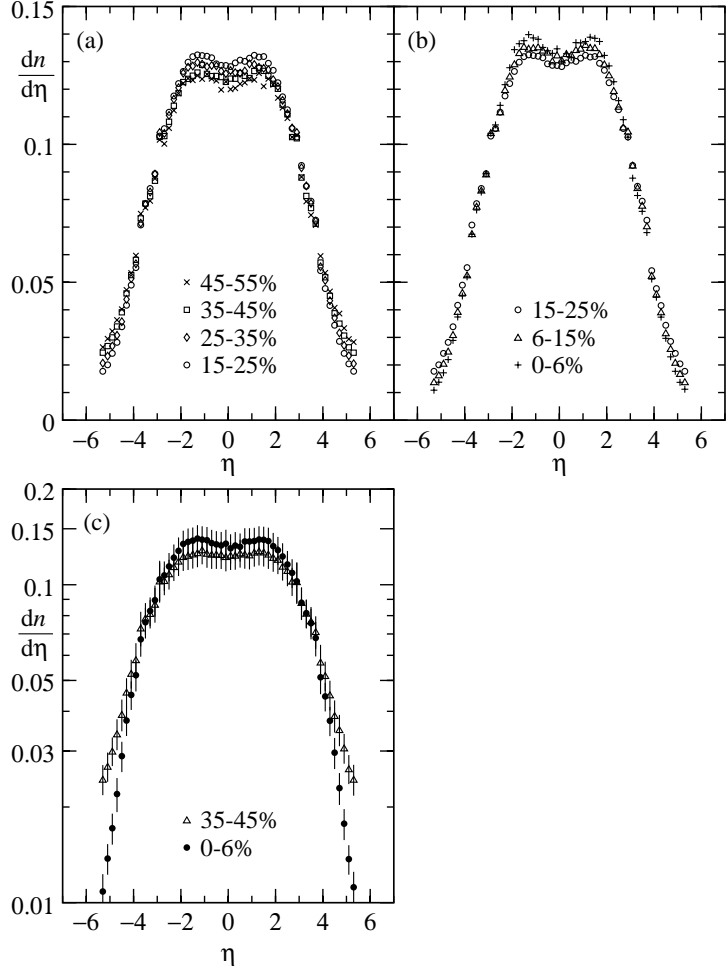}}
  \caption{(a) and (b) Two sets of $dn/d\eta$ with different centrality cuts.\cite{Back:2001bq} Data with 15--25\% are plotted in both figures for the sake of comparison. (c) To examine $dn/d\eta$ in the fragmentation region, a log-linear plot is given.}
\label{fig1}
\end{figure}

In this paper, instead of Eq.~(\ref{eq1.1}), we consider the physical quantity
\begin{eqnarray}
  \frac 1{N_{\rm ch}}\frac{dN_{\rm ch}}{d\eta} = \frac{dn}{d\eta}\:,
\label{eq1.2}
\end{eqnarray}
where $N_{\rm ch} = \int (dN_{\rm ch}/d\eta)d\eta$ and $\int (dn/d\eta)d\eta = 1$. In Fig.~\ref{fig1}, three sets of values for $dn/d\eta$ are shown. They suggest that there is scaling behavior among the different sets of $dn/d\eta$ with different centrality cuts. Thus, $dn/d\eta$ can be considered a kind of probability density, because the variable $\eta$ is a continuous variable. This fact probably implies that a stochastic approach is appropriate in analyses of $dn/d\eta$.\footnote{
From studies of multiparticle dynamics in high energy physics, we have learned that the probability distributions $P(n,\,\langle n\rangle)$ are functions of $n$ and $\langle n\rangle$. It is known that the KNO scaling functions\cite{Koba:1972ng}
$$
\lim_{n,\,\langle n\rangle \to \infty} \langle n\rangle P(n) = \psi (z=n/\langle n\rangle)\:,
$$
are described by solutions of various Fokker-Planck equations.\cite{Biyajima:1984qu,Biyajima:1984qv} Moreover, this stochastic description is also seen in QCD. For example, Dokshitzer has calculated the generalized gamma distribution in QCD\cite{Dokshitzer:1993dc}
$$
P(z=n/\langle n\rangle) \approx \frac{2\mu^2}z \frac{(Dz)^{3\mu/2}}{\sqrt{2\mu\gamma}}\exp \left[-(Dz)^{\mu}\right]\:,
$$
where $z$ is the KNO scaling variable, $\mu$ is given by $1-\gamma = 1/\mu$, $D$ is a parameter, and $\gamma$ is the anomalous dimension in QCD. This is a stationary solution of the following Fokker-Planck equation
$$
\diff Pt = -\diff{}z \left[\left(d+\frac 12Q\right)z + bz^{1+\gamma}\right]P + \frac 12Q\diff{^2}{z^2}[z^2P]\:.
$$
\label{foot2}
}
%
%   FIGURE 1
%

The content of the present paper is as follows. In \S 2, we examine the $dn/d\eta$ scaling. In \S 3, a stochastic approach is considered as one possible explanation of $dn/d\eta$ scaling. In \S 4, concrete analyses employing Gaussian distributions obtained from the Ornstein-Uhlenbeck (O-U) process are presented. In the final section, concluding remarks are given. In the Appendix, the Fokker-Planck equation for the O-U process is considered.

%   SECTION 2
%
\section{Confirmation of \mib{dn/d\eta} scaling}
It is worthwhile to confirm whether $dn/d\eta$ scaling holds. Using the intercepts of $dn/d\eta$ at $\eta = 0$ in Fig.~\ref{fig1},
\begin{eqnarray}
  \left .\frac{dn}{d\eta}\right|_{\eta = 0} = c \approx 0.129\pm 0.005\:,
\label{eq2.1}
\end{eqnarray}
we can obtain a relation between $N_{\rm ch}$ and $\langle N_{\rm part}\rangle$ as follows. The intercept at $\eta=0$ can be parameterized as
\begin{eqnarray}
\left. \frac 1{0.5\langle N_{\rm part}\rangle}\frac{dN_{\rm ch}}{d\eta}\right|_{\eta = 0} = A\langle N_{\rm part}\rangle^{\alpha}\:,
\label{eq2.2}
\end{eqnarray}
where $A=2.16$, and $\alpha = 0.064$. Thus we obtain the relation
\begin{eqnarray}
c = \left. \frac{0.5\langle N_{\rm part}\rangle}{N_{\rm ch}}\frac 1{0.5\langle N_{\rm part}\rangle}\frac{dN_{\rm ch}}{d\eta}\right|_{\eta = 0} = \frac{0.5\langle N_{\rm part}\rangle}{N_{\rm ch}}A\langle N_{\rm part}\rangle^{\alpha}\:.
\label{eq2.3}
\end{eqnarray}
Equation (\ref{eq2.3}) is examined in Table \ref{table1} and Fig.~\ref{fig2}. That Eq.~(\ref{eq2.3}) holds approximately among 6 centrality cuts reflects the $dn/d\eta$ scaling, in particular, in the central region.

%
%   TABLE 1
%
\begin{table}[htb]
\begin{center}
\caption{Empirical examination of Eq.~(\ref{eq2.3}).}
\label{table1}
%\vspace{2mm}
\begin{tabular}{ccc} \hline\hline
centrality (\%) & $2c\times N_{ch}$ & $A\times \langle N_{\rm part}\rangle^{1+\alpha}$ \\ \hline
35--45 & $2\times 0.129 \times 1056$ & $2.16\times 93^{1+0.064}$ \\
 & $272.6\pm 14.6$ & $268.5\pm 17.7$ \\ \hline
25--35 & $2\times 0.129 \times 1582$ & $2.16\times 135^{1+0.064}$ \\
 & $408.1\pm 21.9$ & $399.1\pm 28.4$ \\ \hline
15--25 & $2\times 0.129 \times 2270$ & $2.16\times 197^{1+0.064}$ \\
 & $585.7\pm 31.4$ & $596.7\pm 45.8$ \\ \hline
6--15 & $2\times 0.129 \times 3199$ & $2.16\times 270^{1+0.064}$ \\
 & $825.2\pm 44.6$ & $834.5\pm 67.9$ \\ \hline
0--6 & $2\times 0.129 \times 4070$ & $2.16\times 340^{1+0.064}$ \\
 & $1050\pm  57$ & $1066\pm 90$ \\ \hline
\end{tabular}
\end{center}
\end{table}
%
%   FIGURE 2
%
\begin{figure}
  \epsfxsize= 10 cm   %or \epsfysize= HEIGHT cm
  \centerline{\epsfbox{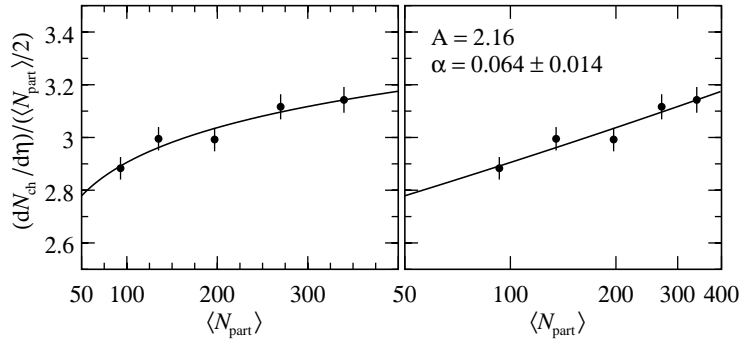}}
  \caption{Determination of the parameters $A$ and $\alpha$.}
\label{fig2}
\end{figure}

%   SECTION 3
%
\section{A possible explanation of \mib{dn/d\eta} obtained using a stochastic approach}
It is well known that the rapidity ($y\approx \eta$) is a kind of velocity. Moreover, there are leading particles in the beam and in target nuclei that collide with each other. For example, nucleons in gold atoms in RHIC experiments collide with each other, and thereby lose energy and emit various particles. Since we have to treat large numbers of particles (e.g., 1k--10k), a stochastic approach seems to be appropriate and to offer a simpler description than Monte Carlo approaches.\footnote{
It should be noted that the transport approach is another useful method (see Refs. \citen{Bass:2001gb} and \citen{Lin:2000cx}).
\label{foot3}
}

To describe $dn/d\eta$ scaling with the leading particle effect and fluctuations in rapidity space, we assume the following Langevin equation \cite{Kampen:1981,Saitou:1980,Hori:1982} for the rapidity variable:\footnote{
In classical mechanics, Eq.~(\ref{eq3.1}) corresponds to the equation
$$
m\frac{dv}{dt} = - m\gamma v + m f_{\rm w}(t)\:,
$$
where $v$ is the velocity.
\label{foot4}
} 
\begin{eqnarray}
  \frac{dy}{dt} = - \gamma y + f_{\rm w}(t)\:.
\label{eq3.1}
\end{eqnarray}
Here $t$, $\gamma$ and $f_{\rm w}(t)$ are the evolution parameter,\footnote{
As an alternative interpretation, $t$ may be related to the number of collisions among the wee partons and produced particles.
\label{foot5}
}
the frictional coefficient and a white noise term, respectively. In our treatment, we assume that $N_{\rm ch}$ particles are produced at $\pm y_{\rm max}$ at $t=0$. This picture takes into account leading particle effects. Using the assumption $y(0) = \pm y_{\rm max}$, we obtain the solution
\begin{eqnarray}
  y(t) = \pm y_{\rm max}e^{-\gamma t} + e^{-\gamma t}\int_0^t e^{\gamma s} f_{\rm w}(s) ds\:.
\label{eq3.2}
\end{eqnarray}
The average and variance of $y(t)$ are calculated as
\begin{eqnarray}
  E[y(t)] = \pm y_{\rm max}e^{-\gamma t}\:,
\label{eq3.3}
\end{eqnarray}
\begin{eqnarray}
  E[ (E[y(t)] - y)^2 ] = \frac{\sigma^2}{2\gamma}\left(1-e^{-2\gamma t} \right)\:,
\label{eq3.4}
\end{eqnarray}
where we use the expression
\begin{eqnarray}
  \langle f_{\rm w}(t)f_{\rm w}(s)\rangle = \sigma^2\delta (t-s)\
\label{eq3.5}
\end{eqnarray}
for the white noise, where $\sigma^2$ is the variance. It is known that the distribution function for $y(t)$ is given by a Gaussian distribution with the above average and variance. The probability density with $V^2(t) = (\sigma^2/2\gamma)(1-e^{-2\gamma t})$ is given by 
\begin{eqnarray}
  P(y,\, y_{\rm max},\, t) &=& 
\frac 1{\sqrt{8\pi V^2(t)}}\left\{
\exp\left[-\frac{(y+y_{\rm max}e^{-\gamma t})^2}{2V^2(t)}\right]\right . 
\nonumber\\
 &&\qquad\qquad\quad\left .+ \exp\left[-\frac{(y-y_{\rm max}e^{-\gamma t})^2}{2V^2(t)}\right]\, \right\}\:.
\label{eq3.6}
\end{eqnarray}
The connection between Eq.~(\ref{eq3.1}) and the Fokker-Planck equation for the O-U process is given in the Appendix. 

\begin{figure}[htb]
\begin{tabular}{cc}
\begin{minipage}{.5\hsize}
  \epsfxsize= 6 cm   %or \epsfysize= HEIGHT cm
  \centerline{\epsfbox{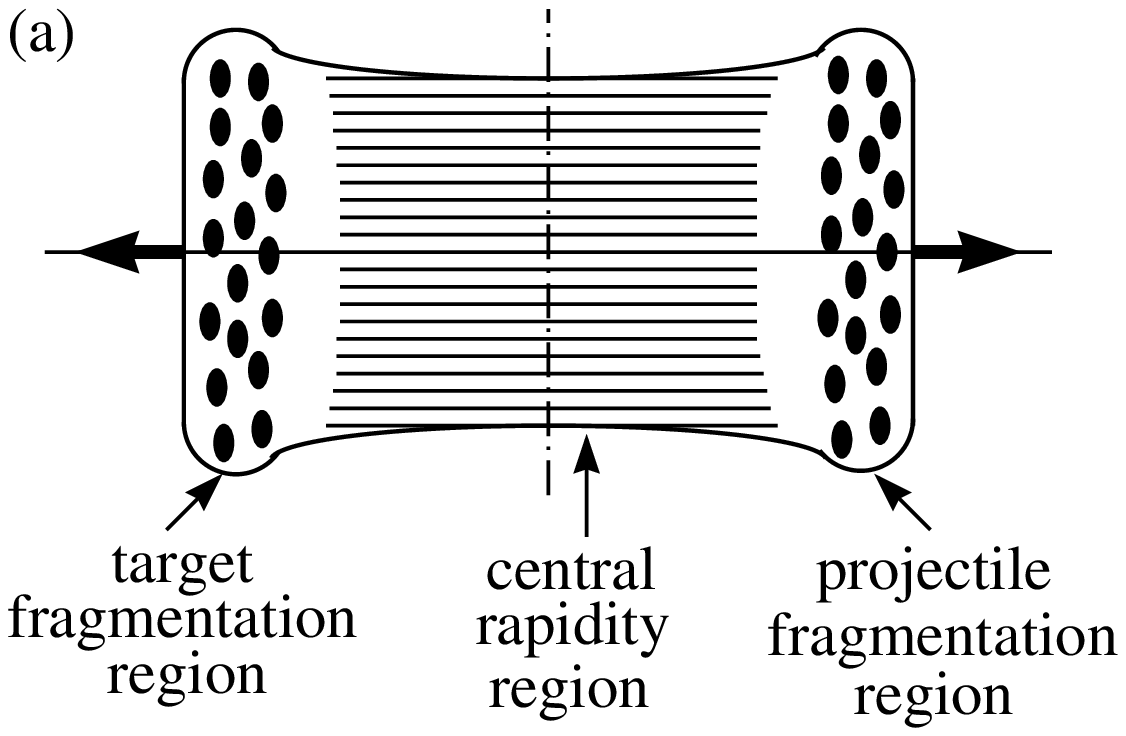}}
\end{minipage}
&
\begin{minipage}{.5\hsize}
  \epsfxsize= 6 cm   %or \epsfysize= HEIGHT cm
  \centerline{\epsfbox{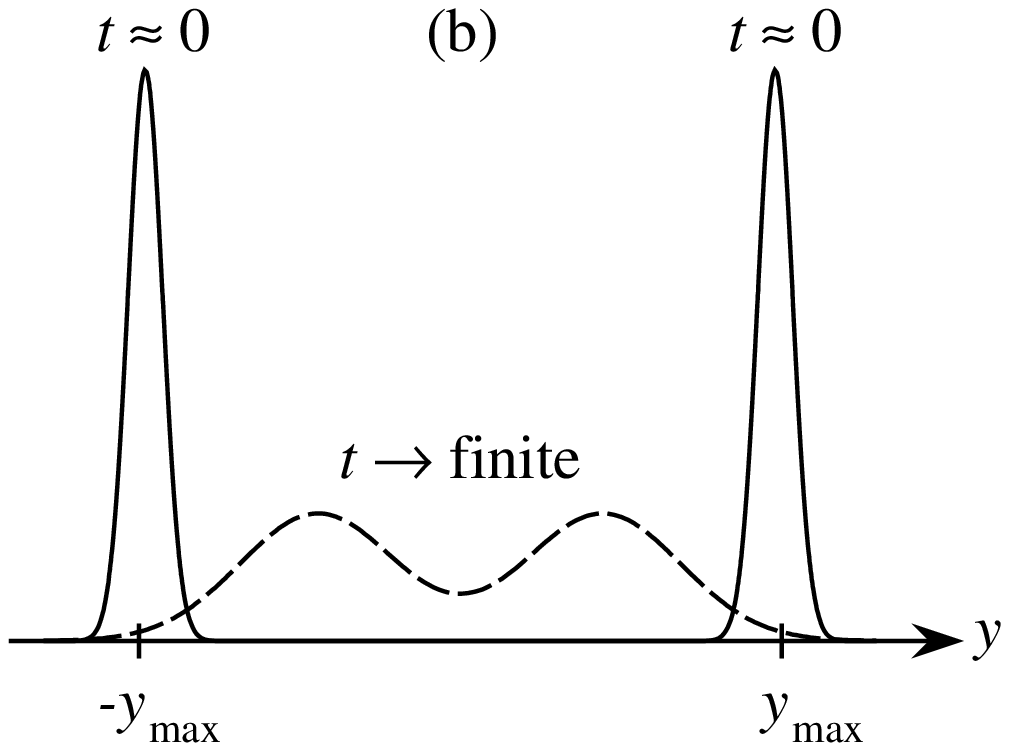}}
\end{minipage}
\end{tabular}
  \caption{(a) Simplified picture for A-A collisions. The thin lines represent partons, and the black circles represent nucleons. (b) Evolution of $P(y,\, y_{\rm max},\, t)$ in Eq.~(\ref{eq3.6}) with two sources at $y_{\rm max}$ and $-y_{\rm max}$.}
\label{fig3}
\end{figure}

In Fig.~3(a), we depict a simplified picture of heavy-ion collision. Our assumptions for the leading particle effect are equivalent to the assumption $P(y,\, y_{\rm max},\, t=0) = 0.5[ \delta (y-y_{\rm max}) + \delta (y+y_{\rm max})]$. In other words, in this model we make the simple assumption that there are two sources of particles at $t=0$, located at $\pm y_{\rm max}$ and producing $0.5 N_{\rm ch}$ particles each. The evolution of $P(y,\, y_{\rm max},\, t)$ given in Eq.~(\ref{eq3.6}) is shown in Fig.~\ref{fig3}(b).
%
%   FIGURE 3
%

%   SECTION 4
%
\section{Analyses of \mib{dn/d\eta} by means of Eq.~(\ref{eq3.6})}
\begin{figure}[htb]
  \epsfxsize= 10 cm   %or \epsfysize= HEIGHT cm
  \centerline{\epsfbox{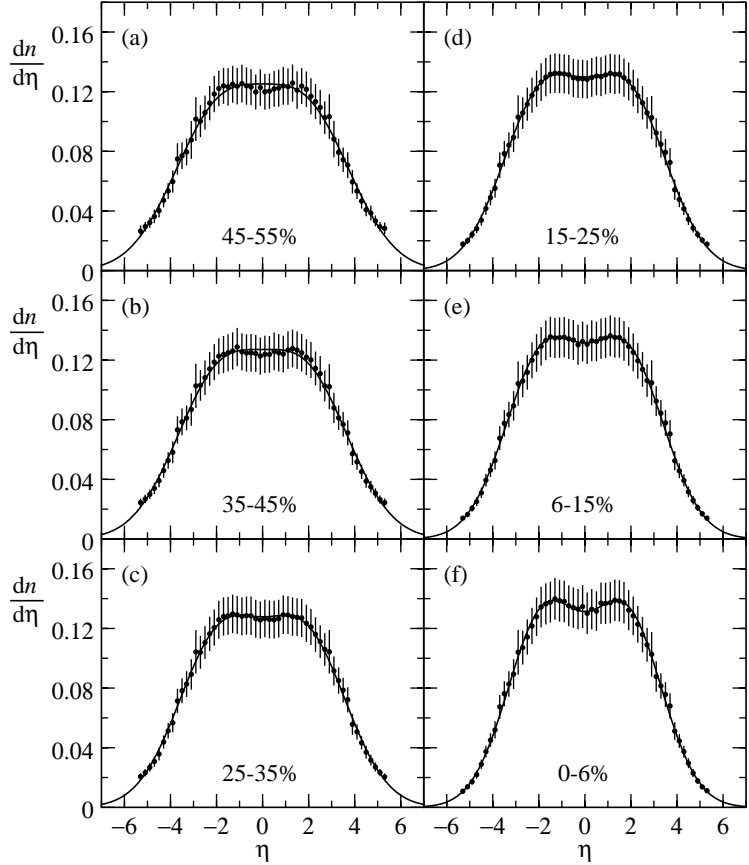}}
  \caption{Analyses of $dn/d\eta$ using Eq.~(\ref{eq3.6}). (See Table \ref{table2}.)}
\label{fig4}
\end{figure}

\begin{figure}[htb]
  \epsfxsize= 7 cm   %or \epsfysize= HEIGHT cm
  \centerline{\epsfbox{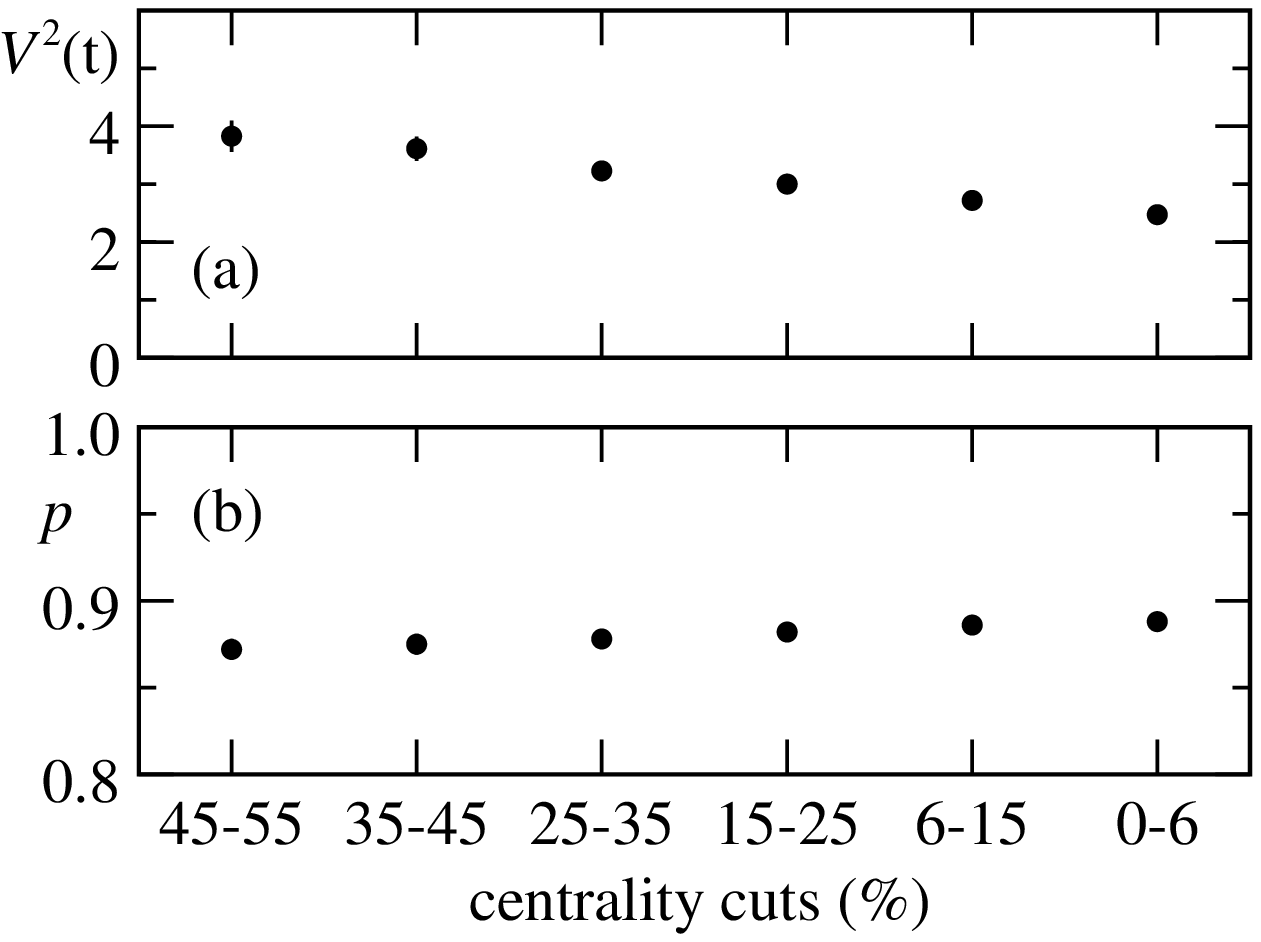}}
  \caption{The values $V^2(t)$ and $p$ of Fig.~\ref{fig4} and Table \ref{table2}.}
\label{fig5}
\end{figure}

Making use of Eq.~(\ref{eq3.6}), we can analyze $dn/d\eta$ shown in Fig.~\ref{fig1}. In our calculation, as most produced particles are not specified, we assume that $y \approx \eta$ in Eq.~(\ref{eq3.6}). Our results are shown in Fig.~\ref{fig4} and Table \ref{table2}. In Fig.~\ref{fig5}, we examine whether or not the variance $V^2(t)$ and the quantity $p=1-e^{-2\gamma t}$ depend on the centrality cuts. As is seen in Figs.~3 and 4, the scaling behavior among the sets of $dn/d\eta$ at $\sqrt{s_{NN}} = 130$ GeV is explained by Eq.~(\ref{eq3.6}), with small changes in the variance $V^2(t)$. The values of $V^2(t)$ depend on the distribution in the fragmentation region [$-\eta_{\rm max} < \eta <-4$ and $4 < \eta < \eta_{\rm max}$]. It can be said that the scaling behavior is explained fairly well by the O-U process with two sources, one at the beam ($y_{\rm B}$ or $y_{\rm max}$) rapidity and one at the target ($y_{\rm T}$ or $-y_{\rm max}$) rapidity. We have confirmed that an O-U process with a single source is not capable of explaining the scaling behavior.\footnote{
For an explanation with the single source at $y_0 \approx 0$, we can use Eq.~(\ref{eqa.2}) in the Appendix. For the centrality cut 0--6\%, we obtain $\chi^2 = 25.76,$ with $m/p_t = 1.3\pm 0.1$, which is necessary for the single source model. If $m/p_t$ is ignored, we obtain a worse value of $\chi^2$. Thus we disregard this model.
\label{foot6}
}$^,$\footnote{
We have investigated whether or not the dip structures at $\eta \approx 0$ can be explained by the Jacobian ${\small p/E}$, and obtained the worse values of $\chi^2$ than those listed in Table \ref{table2}. This fact is probably related to the masses of produced particles that are not measured.
\label{foot7}
}
%
%   FIGURE 4
%
%
%   TABLE 2
%
\begin{table}[htb]
\begin{center}
\caption{Parameter values obtained in our analyses using Eq.~(\ref{eq3.6}) with two sources. The evolution of $P(y,\, y_{\rm max},\, t)$ in Eq.~(\ref{eq3.6}) is stopped at minimum values of $\chi^2$. Here, $\delta p = 0.006--0.004$ and $c= \frac 1{\sqrt{2\pi V^2(t)}}\exp\left[-\frac{(y_{\rm max}e^{-\gamma t})^2}{2V^2(t)}\right]$ . (n.d.f means the number of degree of freedom.)}
\label{table2}
%\vspace{2mm}
\begin{tabular}{ccccccc} \hline\hline
Fig.~\ref{fig4} & (a) & (b) & (c) & (d) & (e) & (f)\\ \hline
centrality (\%) & 45--55 & 35--45 & 25--35 & 15--25 & 6--15 & 0--6 \\
$p$ & 0.872$\pm \delta p$ & 0.875$\pm \delta p$ & 0.878$\pm \delta p$ & 0.882$\pm \delta p$ & 0.886$\pm \delta p$ & 0.888$\pm \delta p$\\
$V^2(t)$ & 3.83$\pm$0.27 & 3.61$\pm$0.21 & 3.23$\pm$0.16 & 3.00$\pm$0.13 & 2.72$\pm$0.10 & 2.47$\pm$0.08\\
$\langle N_{\rm part}\rangle$ & --- & 93 & 135 & 197 & 270 & 340\\
$N_{\rm ch}$ & 662$\pm$10 & 1056$\pm$16 & 1582$\pm$23 & 2270$\pm$34 & 3199$\pm$49 & 4070$\pm$63\\
$c$ & 0.125 & 0.127 & 0.128 & 0.130 & 0.132 & 0.131\\
$\chi^2/{\rm n.d.f.}$ & 8.61/51 & 7.63/51 & 5.88/51 & 5.35/51 & 3.57/51 & 3.82/51\\ \hline
\end{tabular}
\end{center}
\end{table}
%
%   FIGURE 5
%

The intercepts for the sets of $dn/d\eta$ are calculated using the expression
\begin{eqnarray}
c= \frac 1{\sqrt{2\pi V^2(t)}}\exp\left[-\frac{(y_{\rm max}e^{-\gamma t})^2}{2V^2(t)}\right]
\label{eq4.1}
\end{eqnarray}
The results are listed in Table \ref{table2}. They are almost the same as the values in Fig.~\ref{fig1}.

Here we should carefully examine the values of $V(t)$ in Table \ref{table2}. The slight change reflects the discrepancies in the fragmentation region. As seen in Fig.~\ref{fig1}(c), there are small differences in the sets of $dn/d\eta$ for $|\eta| \gsim 4$ between the 0--6\% centrality cut and the other centrality cuts. To explore the differences more carefully, we need sets of $dn/d\eta$ with smaller centrality cuts, for example 0--3\% -- 0--5\%.\footnote{
A calculation based on QCD is given in Ref.~\citen{Kharzeev:2001gp} as
$$
\left .\frac 2{\langle N_{\rm part}\rangle} \frac{dN_{\rm ch}}{d\eta}\right|_{\eta = 0} 
= a\left(\frac s{s_0}\right)^{\lambda/2}\left[\log\left(\frac{Q_{\rm 0S}^2}{\Lambda_{\rm QCD}}\right)+\frac{\lambda}2\log\left(\frac s{s_0}\right)\right]\:,
$$
where $a\approx 0.82$, $\Lambda_{\rm QCD} = 0.2$ GeV and $\lambda = 0.25$, and the centrality dependence of the saturation scale is $Q_{\rm 0S}^2$. In the fragmentation region, this expression needs cutoff factors.
\label{foot8}
}$^,$\footnote{
Very recently, the results for $dn/d\eta$ at $\sqrt{s_{NN}} = 130$ GeV and $200$ GeV with centrality cuts 0--5\% and others obtained by the BRAHMS Collaboration\cite{Bearden:2001xw,Bearden:2001qq} were reported. We have analyzed them using Eq.~(\ref{eq3.6}) and obtained an almost constant value of $V^2(t)$, because data in the fragmentation region are lacking for $|\eta| \gsim 4.5$.
\label{foot9}
}

%   SECTION 5
%
\section{Concluding remarks}
First, it can be said that there is scaling among the different sets of $dn/d\eta$ with various centrality cuts at $\sqrt{s_{NN}} = 130$ GeV, as seen from the nearly constant values of $c$ and the behavior in Figs.~\ref{fig1}(a)--(c). 

Second, the scaling behavior of $dn/d\eta$ is described by the solution given in Eq.~(\ref{eq3.6}) of the Langevin equation with two sources.\footnote{
To estimate the ``thermalization time'' of the QGP, Hwa has considered the Fokker-Planck equation for the motion of quarks and gluons in nuclei.\cite{Hwa:1985fj} (See also Ref.~\citen{Chakraborty:1984ha}, in which the Wiener process is considered for the problem of the thermalization of quarks and gluons.)
\label{foot10}
}(See Fig.~\ref{fig4} and the values of $c$ in Table \ref{table2}.)

\begin{figure}[htb]
  \epsfxsize= 7 cm   %or \epsfysize= HEIGHT cm
  \centerline{\epsfbox{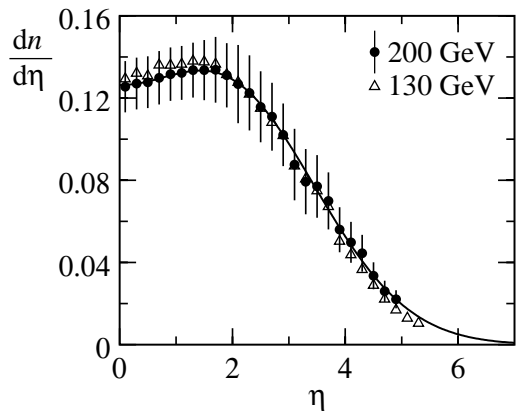}}
  \caption{Comparisons of $dn/d\eta$ with the centrality cut 0--6\% at $\sqrt{s_{NN}} = 130$ GeV and 200 GeV. The solid curve is obtained for latter energy. Here $p = 0.879\pm 0.007$, $V^2(t) = 2.67\pm 0.24$, $\chi^2/{\rm n.d.f.} = 0.63/22$ and $c=0.126$.}
\label{fig6}
\end{figure}

Third, we can add the following fact. Very recently, data for $dN_{\rm ch}/d\eta$ with centrality cut 0--6\% at $\sqrt{s_{NN}} = 200$ GeV from the PHOBOS Collaboration were reported.\cite{Back:2001ae}  They are compared with data for $dn/d\eta$ at $\sqrt{s_{NN}} = 130$ GeV in Fig.~\ref{fig6}. From this comparison, it is obvious that the scaling of $dn/d\eta$ holds between $\sqrt{s_{NN}} = 130$ GeV and 200 GeV.\footnote{
Using Bjorken's picture \cite{Bjorken:1983qr} for the calculation of the energy density for $|\Delta \eta|\leq 0.5$ with a geometrical picture of gold ($R_{\tau} \approx$ 6--7 fm, $c\tau_0 \approx$ 1--2 fm, $V \approx \pi R_{\rm T}^2(c\tau_0) \approx 300\ {\rm fm^3}$), we obtain the following values:
\begin{eqnarray*}
  \varepsilon &\sim& \left. \frac 32 \frac 1V \frac{dN_{\rm ch}}{d\eta}E_{\rm T} \right|_{|\Delta\eta|\leq 0.5} \sim 1\ {\rm GeV/fm^3}\quad (130\ {\rm GeV})\, ,\\
  \varepsilon &\sim& 1.2\ {\rm GeV/fm^3}\quad (200\ {\rm GeV})\:.
\end{eqnarray*}
\label{foot11}
}
%
%   FIGURE 6
%

Moreover, we can consider the $dn/d\eta$ scaling from a different point of view, i.e., regarding the scaling property of Gaussian distributions. Using $\eta_{\rm rms} = \sqrt{\langle \eta^2\rangle} = \sqrt{\sum \eta^2dn/d\eta}$, we can compute the following quantity with $z_r = \eta/\eta_{\rm rms}$ :
\begin{eqnarray}
\eta_{\rm rms}\frac{dn}{d\eta} = \frac{dn}{dz_r} = f(z_r=\eta/\eta_{\rm rms})\:.
\label{eq5.1}
\end{eqnarray}
In Fig.~\ref{fig7}, we display our result for Eq.~(\ref{eq5.1}) and give $\eta_{\rm rms}$ in Table \ref{table3}. This scaling of $f(z_r=\eta/\eta_{\rm rms})$ reflects the fact that the sets of $dn/d\eta$ are described by a Gaussian distribution. In other words, the description of $dn/d\eta$ using the O-U process is appropriate.
%
%   FIGURE 7
%
\begin{figure}[htb]
  \epsfxsize= 8 cm   %or \epsfysize= HEIGHT cm
  \centerline{\epsfbox{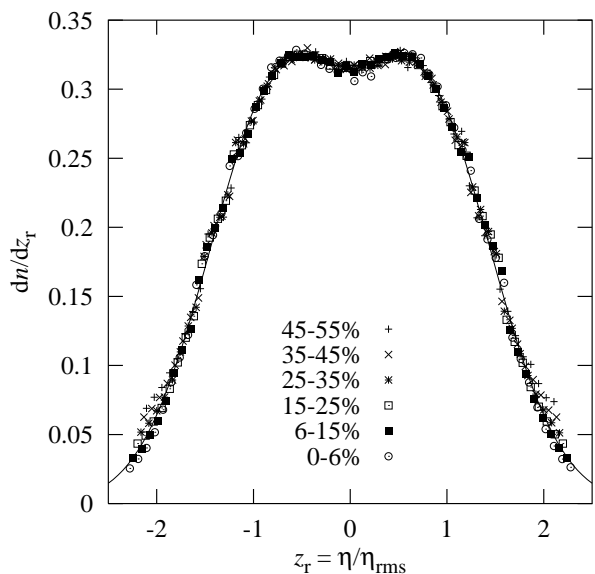}}
  \caption{Normalized distribution of $dn/dz_r$ with $z_r=\eta/\eta_{\rm rms}$ scaling. The solid curve is obtained by $\frac{dn}{dz_r} = \frac 1{\sqrt{8\pi V_r^2(t)}}\left\{\exp\left[-\frac{(z_r+z_{\rm max}e^{-\gamma t})^2}{2V_r^2(t)}\right]+ \exp\left[-\frac{(z_r-z_{\rm max}e^{-\gamma t})^2}{2V_t^2(t)}\right]\, \right\}$, $z_{\rm max} = 2.23$, $V_r^2(t)=0.502$, $\chi^2/{\rm n.d.f.} = 92.9/321$.}
\label{fig7}
\end{figure}
%
%   TABLE 3
%
\begin{table}[htb]
\begin{center}
\caption{Values of $\eta_{\rm rms}$ at $\sqrt{s_{NN}} = 130$ GeV}
\label{table3}
%\vspace{2mm}
\begin{tabular}{ccccccc} \hline\hline
Fig.~\ref{fig4} & (a) & (b) & (c) & (d) & (e) & (f)\\ \hline
centrality (\%) & 45--55 & 35--45 & 25--35 & 15--25 & 6--15 & 0--6 \\
$\eta_{\rm rms} = \sqrt{\langle \eta^2\rangle}$ & 2.52 & 2.49 & 2.45 & 2.41 & 2.37 & 2.33\\ \hline
\end{tabular}
\end{center}
\end{table}

This $dn/d\eta$ scaling suggests that $dn/d\eta$ with the centrality cut 0--6\% does not exhibit singular or particular phenomena related to signatures of the quark-gluon plasma (QGP).\footnote{
From studies of the HBT effect at RHIC, the authors of Ref.~\citen{Adler:2001nb} have concluded that the expected large radius from the QGP is not observed.
\label{foot12}
}Of course, care must be taken when handling averaged quantities in statistics. Thus, while at present, it appears that the QGP is not created, it is possible that the QGP is created but its signature is washed out by strong interactions between hadrons.\footnote{
Through the measurement of $v_2$ (flow of nucleons), some physicists have conjectured that the increasing value of $v_2$ reflects the effect from the compression of nuclear matter \cite{Ackermann:2000tr}. (See also Ref.~\citen{Hirano:2001eu}.)
\label{foot13}
}

To investigate particular phenomena, like the turbulence and/or deflagration in $dN_{\rm ch}/d\eta$, we need to analyze a single event with smaller centrality cut than 0--6\%.

Furthermore, event-by-event analyses using intermittency\cite{Burnett:1983pb,Takagi:1984gr,Bialas:1986jb,Biyajima:1990kw,Andreev:1995rc,Hwa:1990xg} and wavelets\cite{Suzuki:1995kg} are necessary to investigate the detection of QGP and a disoriented chiral condensate~(DCC). For the latter case, the ratio of neutral pions ($\langle \pi^0\rangle$) to charged pions ($\langle \pi^{\rm ch}\rangle$), $\langle \pi^0\rangle/\langle \pi^{\rm ch}\rangle$, should be measured. These methods should be applied to $dN_{\rm ch}/d\eta$ with smaller centrality cuts and larger particles. They seem to be capable of extracting useful information on QGPs and DCCs from the analysis of single events.

Finally, we should mention the results of recent analyses of data concerning $dN_{\rm ch}/d\eta$ carried out by the NA50 Collaboration.\cite{Abreu:2002fx} It has been reported that there is $\eta$-scaling of $dN_{\rm ch}/d\eta$ in the data obtained from Pb + Pb collisions at SPS.\cite{Abreu:2002fx} This behavior can be described by a single Gaussian distribution. (See Eq.~(\ref{eqa.2}).) This fact suggests that the stochastic approach can be used to describe the behavior of $dN_{\rm ch}/d\eta$ at SPS. The intercepts at $\eta_{\rm max}$, $(N_{\rm max})^{-1}dN_{\rm ch}/d\eta = dn/d\eta$ are in range 0.246--0.266 for 6 centrality cuts. It should be noticed that there is no dip structure. Their observation supports the validity of our stochastic approach.
\smallskip\\
{\it Note added in proof:} After completion of this paper, we are informed that the relativistic diffusion model has been used for analysis of the proton distributions in heavy ion collisions by Wolschin.\cite{Wolschin:1999jy} A similar Fokker-Planck equation is used therein.

%   ACKNOWLEDGEMENT
%
\section*{Acknowledgements}
One of authors (M. B.) is partially supported by a Grant-in-Aid from the Ministry of Education, Science, Sports and Culture, Japan, (No. 09440103). Useful conversations with G. Wilk, T. Osada and S. Raha are gratefully acknowledged.

%   APPENDIX
%
\appendix
\section{Solution of the Fokker-Planck Equation for the O-U Process} %Empty argument \section{} yields `Appendix'. 
The Fokker-Planck equation for the O-U process connected with Eq.~(\ref{eq3.1}) is given by
\begin{eqnarray}
  \diff{P(y,\, t)}t = \gamma \left[\diff{}yy + \frac 12\frac{\sigma^2}{\gamma}\diff{^2}{y^2}\right] P(y,\, t)\:.
\label{eqa.1}
\end{eqnarray}
The solution of Eq.~(\ref{eqa.1}) with $P(y,\, 0) = \delta (y - y_0)$ is obtained as 
\begin{eqnarray}
  P(y,\, t) = \frac 1{\sqrt{2\pi V^2(t)}}\exp\left[-\frac{(y-y_0e^{-\gamma t})^2}{2V^2(t)}\right]\:,
\label{eqa.2}
\end{eqnarray}
where we have used a resolution method for partial differential equations, including the equation of the characteristic.\cite{Goel:1974} Similarly, the solution of Eq.~(\ref{eqa.1}) with $P(y,\, 0) = 0.5[\delta (y + y_{\rm max})+\delta (y - y_{\rm max})]$ is obtained as
\begin{eqnarray}
  P(y,\, y_{\rm max},\, t) &=& 
\frac 1{\sqrt{8\pi V^2(t)}}\left\{
\exp\left[-\frac{(y+y_{\rm max}e^{-\gamma t})^2}{2V^2(t)}\right]\right . 
\nonumber\\
 &&\qquad\qquad\quad\left .+ \exp\left[-\frac{(y-y_{\rm max}e^{-\gamma t})^2}{2V^2(t)}\right]\, \right\}\:.
\label{eqa.3}
\end{eqnarray}

%   REFERENCES
%

\newpage
\section*{Addenda} %Empty 
\setcounter{section}{2}
\setcounter{equation}{0}

\begin{center}
{\large Scaling Behavior of $(N_{\rm ch})^{-1}dN_{\rm ch}/d\eta$ at $\sqrt{s_{NN}} = 130\ {\rm GeV}$ by the PHOBOS Collaboration and Its Implication}\medskip\\
--- A Possible Explanation Employing the Ornstein-Uhlenbeck Process ---\medskip\\
Minoru Biyajima, Masaru Ide, Takuya Mizoguchi and Naomichi Suzuki\medskip\\
Prog.~Theor.~Phys.\ {\bf 108} (2002), 559-569.
\end{center}

\vspace{7mm}

Very recently PHOBOS Collaboration has reported new analysis of $d N_{ch}/d\eta$ at $\sqrt{s_{NN}} = 130$ GeV.\cite{Nouicer:2002ks} Thus we should add several new results based on the stochastic theory using their new data. 

Before our new analyses, the following re-calculations are necessary in Fig.~\ref{fig2} and Table~\ref{table1}, because an empirical value at $\langle N_{\rm part}\rangle = 93$ and the magnitudes of error bar should be corrected. The method of the linear regression is used, because values of $\langle N_{\rm part}\rangle$ are computed by the Glauber model. There is no change in our concluding remarks, since Eq.~(2.1) is explained by Eq.~(2.3).
%
%   FIGURE 2
%
\begin{figure}[htb]
\begin{tabular}{cc}
\begin{minipage}{.47\hsize}
  \setcounter{figure}{1}
  \epsfxsize= 5.5 cm   %or \epsfysize= HEIGHT cm
  \centerline{\epsfbox{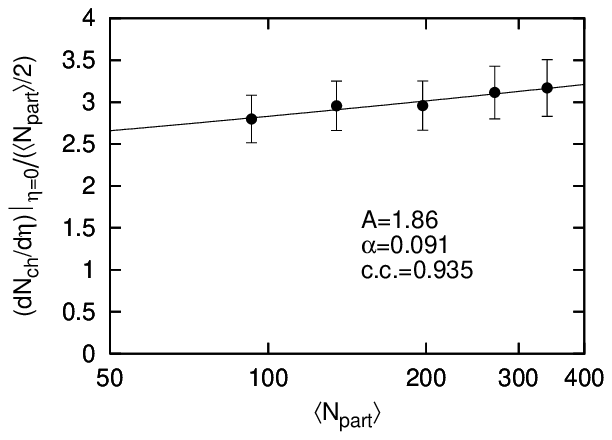}}
  \caption{Parameters A and $\alpha$ are determined by the method of linear-regression. The correlation coefficient (c.c.) is 0.935.}
\label{fig2}
\end{minipage}
&
\begin{minipage}{.47\hsize}
\setcounter{figure}{7}
  \epsfxsize= 5.5 cm   %or \epsfysize= HEIGHT cm
  \centerline{\epsfbox{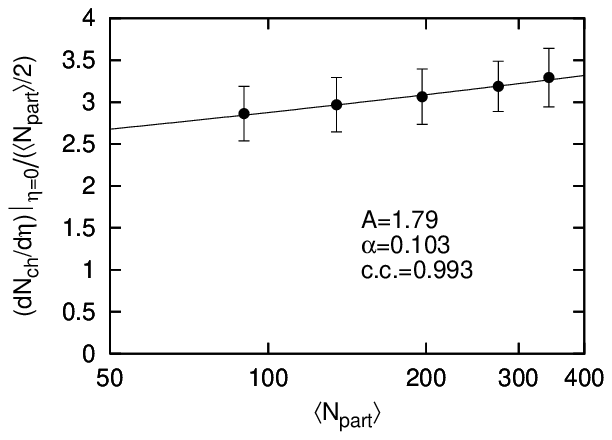}}
  \caption{Same as Fig.2, this calculation is done by new data.\cite{Nouicer:2002ks} The correlation coefficient (c.c.) is 0.993.}
\label{fig8}
\end{minipage}
\end{tabular}
\end{figure}
%
%   TABLE 1
%
\setcounter{table}{0}
\begin{table}[htb]
\begin{center}
\caption{Empirical examination of Eq.(2.3).}
\label{table1}
%\vspace{2mm}
\begin{tabular}{ccccccc} \hline\hline
centrality (\%) & 35--45 & 25--35 & 15-25 & 6--15 & 0--6\\ \hline
$\langle N_{\rm part}\rangle$ & $93$ & $135$ & $197$ & $270$ & $340$\\ \hline
$N_{ch}$ & $1030\pm 110$ & $1550\pm 160$ & $3160\pm 330$ & $2230\pm 230$ & $4030\pm 430$\\ \hline
c (Eq.~(2.3)) & $0.127\pm 0.014$ & $0.127\pm 0.013$ & $0.133\pm 0.014$ & $0.132\pm 0.01$ & $0.134\pm 0.014$\\ \hline
\end{tabular}
\end{center}
\end{table}

Next we have to examine Eqs.~(2.1)-(2.3), using new data in Ref.~\citen{Nouicer:2002ks}. The intercept of empirical values with 6 centrality cuts is $dn/d\eta|_{\eta = 0} = c \approx 0.132 \pm 0.009$. For Eq.~(2.2), we obtain the parameters $A=1.79$ and $\alpha = 0.103$. Eq. (2.3) becomes as follows
\begin{eqnarray*}
c = \frac{0.5\langle N_{\rm part}\rangle}{N_{\rm ch}}\cdot 1.79\langle N_{\rm part}\rangle^{0.103}\:.
\label{eq6}
\end{eqnarray*}
See Fig.~\ref{fig8} and Table~\ref{table4}. Because of larger c.c., we can conclude that Eq.~(2.3) holds for new data of Ref.1), i.e., $c = 0.132 \pm 0.005$. 

\begin{figure}[tbh]
\begin{tabular}{cc}
\begin{minipage}{.47\hsize}
%
%   FIGURE 9
%
  \epsfxsize= 7 cm   %or \epsfysize= HEIGHT cm
  \centerline{\epsfbox{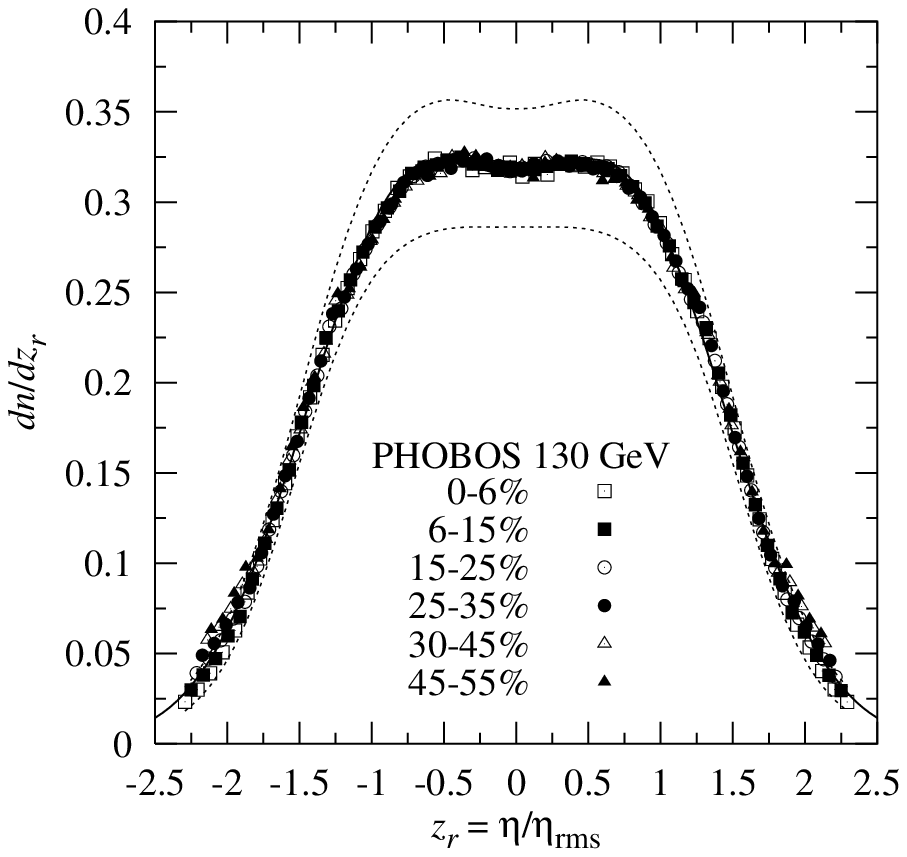}}
\end{minipage}
&
\begin{minipage}{.47\hsize}
  \caption{${\rm z}_r$ scaling at $\sqrt{s_{NN}}$=130 GeV. The dashed lines represent of error bars with the centrality cut 0-6$\%$. The solid line is obtained by Eq.~(\ref{eqb1}) with $z_{\rm max} = 2.03$, $p = 0.854\pm 0.002$, $V_r^2(t) = 0.49\pm 0.01$ and $\chi^2/{\rm n.d.f.} = 25.5/321$.}
\label{fig9}
\end{minipage}
\end{tabular}
\end{figure}

Third, we apply Eq.~(3.6) to $d n/d\eta$ distribution and obtain estimated values given in Table~\ref{table5}.
%
%   TABLE 4
%
\setcounter{table}{3}
\begin{table}[htb]
\begin{center}
\caption{Empirical examination of Eq.(2.3) in new data.\cite{Nouicer:2002ks}}
\label{table4}
%\vspace{2mm}
\begin{tabular}{ccccccc} \hline\hline
centrality (\%) & 35--45 & 25--35 & 15-25 & 6--15 & 0--6\\ \hline
$\langle N_{\rm part}\rangle$ & $90 \pm 5$ & $135 \pm 6.0$ & $196.5 \pm 7.5$ & $274.5 \pm 8.5$ & $342.5 \pm 11.0$\\ \hline
$N_{ch}$ & $1000\pm 50$ & $1540\pm 80$ & $2270\pm 110$ & $3230\pm 160$ & $4100\pm 210$\\ \hline
c (Eq.~(2.3)) & $ 0.128\pm 0.010$ & $0.130\pm 0.009$ & $0.133\pm 0.009$ & $0.136\pm 0.008$ & $0.136\pm 0.008$\\ \hline
\end{tabular}
\end{center}
\end{table}
%
%   TABLE 5
%
%\vspace{-2mm}
\begin{table}[htb]
\begin{center}
\caption{New results from analysis of $dN_{ch}/d\eta$ at $\sqrt{s_{NN}}=130$ Gev in Ref.~1) by means of Eq.~(3.6), $c({\rm Th}) = \left(1/\sqrt{2\pi V^2(t)}\right)\exp\left[-(\eta_{\rm max}e^{-\gamma t})^2/(2V^2(t))\right]$ and $\eta_{max}=4.9$. Here, $\delta p =$ 0.004--0.006 and $\delta c =$ 0.005--0.006. (n.d.f. means the number of degree of freedom.)}
\label{table5}
%\vspace{2mm}
\begin{tabular}{ccccccc} \hline\hline
centrality (\%) & 45--55 & 35--45 & 25--35 & 15-25 & 6--15 & 0--6  \\ \hline
$\eta_{rms}$  & 2.51 & 2.48 & 2.44 & 2.40 & 2.36 & 2.31 \\ \hline
$p$  & $0.839\pm \delta p$ & $0.845\pm \delta p$ & $0.850\pm \delta p$ & $0.857\pm \delta p$ & $0.862\pm \delta p$ & $0.866\pm \delta p$ \\
$V^2(t)$ & $3.46\pm 0.25$ & $3.33\pm 0.23$ & $3.07\pm 0.18$ & $2.87\pm 0.15$ & $2.66\pm 0.12$ & $2.46\pm 0.11$ \\
$c$(Th) & $0.123\pm \delta c_t$ & $0.125\pm \delta c_t$ & $0.127\pm \delta c_t$ & $0.130\pm \delta c_t$ & $0.131\pm \delta c_t$ & $0.132\pm \delta c_t$ \\
$\chi^2/{\rm n.d.f.}$  & 2.3/51 & 2.3/51 & 2.2/51 & 1.6/51 & 2.0/51 & 2.7/51 \\ \hline
\end{tabular}
\end{center}
\end{table}
For the $z_r = \eta/\eta_{\rm rms}$ scaling, Eq.~(5.1) and the following Eq.~(\ref{eqb1}) are applied to data of Ref.~\citen{Nouicer:2002ks}. The distribution by $z_r$ scaling variable is shown in Fig.~\ref{fig9}. Theoretical formula is given as
\begin{eqnarray}
  \frac{dn}{dz_r} &=& \frac{1}{\sqrt{8\pi V_r^2(t)}}\left\{\exp\left[-\frac{(z_r+z_{\rm max}e^{-\gamma t})^2}{2V_r^2(t)}\right]+ \exp\left[-\frac{(z_r-z_{\rm max}e^{-\gamma t})^2}{2V_r^2(t)}\right]\, \right\}\nonumber\\
&&\label{eqb1}
\end{eqnarray}
with $z_{\rm max} = \eta_{\rm max}/\eta_{\rm rms}$ and $V_r^2(t) = V^2(t)/\eta_{\rm rms}^2$. The denominator of $z_{\rm max}$ is an averaged $\eta_{rms}$ among 6 centrality cuts. Figure~\ref{fig9} shows better $z_r$ scaling than the distribution of Fig.~7. It can be stressed that this $z_r$ scaling seems to be useful to compare $d n/d\eta$'s at different energies.\cite{Biyajima:2002wq}
\\
{\it Acknowledgements}\quad Authors would like to thank R.~Nouicer of PHOBOS Collaboration for useful communications. 

\end{document}